\documentclass[preprint2]{aastex631}
\usepackage{natbib}
\usepackage{graphicx}
\usepackage{wasysym}
\usepackage{txfonts}
\usepackage{ulem}
\def\fH2{\mbox {f$_{{\rm H}_2}$}}

\def\EBV{\mbox{\rm{E(B-V)}}}

\def\AV{\mbox{A$_{\rm V}$}}

\def\nH2{\mbox{${\rm n}(\HH$)}}
\def\enH2{\mbox{$n_{(\HH$)}}}

\def\pccc{~{\rm cm}^{-3}} 
 
\def\pcc {\mbox{${~{\rm cm}^{-2}}$}}

\def\Tsub#1 {\mbox{${\rm T}_{\rm #1}$}}
\def\TK  {\Tsub K }

\def\arcsec{\mbox{$^{\prime\prime}$}} \def\arcmin{\mbox{$^{\prime}$}}

\def\degr{\mbox{$^{\rm o}$}}
\def\p{\mbox{$^+$}}
\def\z{\mbox{${^0}$}}

\def\cch{\mbox{C$_2$H}}

\def\h13cop{\mbox{{H$^{13}$CO\p}}}

\def\c3h2{\mbox{C$_3$H$_2$}}

\def\cyclic{\mbox{$c$-C$_3$H$_2$}}
 
 \def\R0{R$_0$}
\def\G0{\mbox{G$_0$}}

\def\ddeg{{}^\circ\kern-.1em}

\def\kms{\mbox{km\,s$^{-1}$}}

\def\m1{\mbox{$^{-1}$}}

\def\E#1 {$10^{#1}$}
\def\E#1 {E{#1}}
\def\P#1,{$\nH2\TK~=~#1\times~10^4\pccc$~K}
\def\ec#1,#2,#3,{#1\,(#2)\E{#3}}

\def\H3{\mbox{H$_3$}}
\def\Lya{\mbox{Lyman-$\alpha$}}

\def\RH2{\mbox{R$_{\rm G}$}}
\def\g13{\mbox{g$_{13}$}}

\def\kHeH2{\mbox{$k_{ He-\HH}$}}

\def\tim#1,#2{\mbox{{$#1\times10^{#2}$}}}

\def\Whcop{\mbox{$\Upsilon_{{\rm HCO}\p}$}}

\def\L21{\mbox{{$\lambda$21cm}}}

\newcommand{\emm}[1]{\ensuremath{#1}}   % ensures math mode.
\newcommand{\emr}[1]{\emm{\mathrm{#1}}} % uses math roman fonts.
\newcommand{\hcop}{\emr{HCO^+}} 
 
\newcommand{\HH}{\emr{H_2}}
\newcommand{\cotw}{\emr{^{12}CO}}
\renewcommand{\coth}{\emr{^{13}CO}}

\newcommand{\N}[1]{\emr{N_{#1}}}

\newcommand{\X}[1]{\emm{X_\emr{#1}}}

\newcommand{\W}[1]{\emm{{\rm W}_\emr{#1}}}
\newcommand{\WCO}{\W{CO}}

\newcommand{\f}[1]{\emm{f_\emr{#1}}}

\shorttitle{Hydrogen, \HH\ and their proxies}
\shortauthors{Harvey Liszt \& Maryvonne Gerin}

\begin{document}


\title{Molecular hydrogen and its proxies HCO$^+$ and CO in the diffuse interstellar medium}


\author{Harvey Liszt}
\affil{National Radio Astronomy Observatory \\
        520 Edgemont Road, Charlottesville, VA 22903\\
        hliszt@nrao.edu}
\author{Maryvonne Gerin}
\affil{LERMA, Observatoire de Paris, PSL Research University, \\
 CNRS, Sorbonne Universit\'e \\
 maryvonne.gerin@observatoiredeparis.psl.eu}
%\and
%\author{Isabelle Grenier}
%\affil{Laboratoire AIM, CEA-IRFU/CNRS/Universit\'e de Paris, D\'epartement d'Astrophysique, \\
%CEA Saclay, F-91191 Gif sur Yvette, France \\
%isabelle.grenier@cea.fr}

\email{hliszt@nrao.edu}

%\today



\begin{abstract}

There is a robust polyatomic chemistry in diffuse, partially-molecular interstellar 
gas that is readily accessible in absorption at radio/mm/sub-mm wavelengths.
Accurate column densities are derived owing to the weak internal excitation, 
so relative molecular abundances are well known with respect to each other but 
not with respect to H$_2$. Here we consider the use of proxies for hydrogen 
column densities N(H$_2$) and N(H) = N(HI)+2N(H$_2$) based on measurements of 
HCO$^+$ absorption and CO emission and absorption, and we compare these with 
results obtained by others when observing HI, H$_2$ and CO toward stars and 
AGN.  We consider the use
of HCO$^+$ as a proxy for H$_2$ and show that the assumption of a relative 
abundance N(H$_2$) = N(HCO$^+$)$/3\times10^{-9}$ gives the same view of the 
atomic-molecular hydrogen transition that is seen in UV absorption toward stars. 
CO on the other hand shows differences between the radio and optical regimes
because emission is always detected when N(\hcop) $\ga 6\times 10^{11}\pcc$
or N(H$_2$) $\ga 2\times 10^{20}\pcc$.
%, implying N(CO) $\ga 10^{14}\pcc$. 
Wide variations in the integrated CO {J=1-0} brightness W$_{\rm CO}$ 
and N(CO)/N(H$_2$)  
imply equivalent variations in the CO-H$_2$ conversion factor 
%N(H$_2$)/W$_{\rm CO}$ 
even while the ensemble mean is near the usual Galactic values.
Gas/reddening ratios found in absorption toward stars, 
N(H)/E(B-V) = 
$6.2 \times 10^{21}~{\rm H} \pcc$ (mag)$^{-1}$ overall or 
$6.8 \times 10^{21}~{\rm H} \pcc$ (mag)$^{-1}$ for sightlines
at E(B-V) $\leq$ 0.08 mag lacking H$_2$ are  well below the 
Galactic mean measured at low reddening and high Galactic latitude,
$8.3 \times 10^{21}~{\rm H} \pcc$ (mag)$^{-1}$.
  
 \end{abstract}



%s1
\section{Introduction}

The discovery of \cyclic\ and \hcop\ at radio frequencies \citep{CoxGue+88,LucLis93} 
pointed the way to an unexpectedly accessible and complex absorption line chemistry 
of diffuse interstellar gas seen along sightlines having extinction \AV\ $\leq 1$ mag.  
The molecular inventory now includes more than three dozen species as complex as 
\HH CO, NH$_3$ \citep{Nas90,LisLuc+06} and CH$_3$CN \citep{LisGer+18a}. There is
a wide range neutral hydrides and hydride ions \citep{GerNeu+16}, linear and cyclic hydrocarbons 
CH, \cch, C$_3$H and C$_3$\HH\ \citep{LucLis00C2H,LisSon+12,LisGer+18a}, sulfur 
compounds CS, SO, \HH S, SH and HCS\p \citep{LucLis02,NeuGod+15}.  There is a diverse group of 
molecular ions CH\p, HOC\p, CF\p\ and C$_3$H\p, \citep{GerLis+19} in addition to 
\hcop\ whose use as a tracer of \HH\ is the subject of this work. A recent search for 
CO\p\ was unsuccessful \citep{GerLis21}.

Half of the diatomics detected in diffuse gas are observed in UV/optical 
absorption but among the polyatomics only a few are observed outside the radio 
waveband, including C$_3$ in the optical \citep{MaiLak+01} and H$_3$\p\ in the 
near IR \citep{GebOka96}. Only CO, OH, and CH are widely observable in emission.  
For this reason the range of the diffuse gas chemistry is accessible only in 
absorption on sightlines toward radio continuum sources that are widely 
separated over the sky and only seldom are bright enough to observe 
spectroscopically at both optical and radio wavelengths \citep{Lis21Tappe}.

This has the consequence that the abundances of the molecules observed at 
radio wavelengths are well-determined and known with respect to each other 
but not directly known with respect to molecular hydrogen. 
However, CH, OH and CO are observed in UV/optical absorption toward stars 
along with \HH, and they are also observable at radio/sub-mm frequencies. 
The relative abundances 
X(CH) = N(CH)/N(\HH) $= 3.5\times 10^{-8}$ \citep{SheRog+08,WesGal+10} and 
X(OH) = N(OH)/N(\HH) $= 10^{-7}$ \citep{WesGal+10} are observed to be fixed 
in UV/optical absorption, with the consequence that the relative abundance 
X(\hcop) = N(\hcop)/N(\HH) $= 3.0\times 10^{-9}$ is
also known with a logarithmic uncertainty of 0.2 dex \citep{GerLis+19}.

The existence of a trusted \hcop/\HH\ ratio puts the polyatomic diffuse 
cloud chemistry on a firmer basis and in two recent works we explored
the use of \hcop\ to assess the molecular component of the dark neutral 
medium (DNM) \citep{Pla15Cham,RemGre+18} in the outskirts of the Chamaeleon 
clouds \citep{LisGer+18}. We found that the DNM was mostly molecular even
while the medium as a whole was predominantly atomic.  We explained the 
absence of CO emission associated with the DNM as the result of very low 
CO/\HH\ relative abundances as determined in CO and \hcop\ absorption
\citep{LisGer+19}. 

We have since acquired many more ALMA and IRAM 30m measurements of \hcop\ 
and CO along directions in the outer Galaxy and in forthcoming work we will 
use that data to study the DNM and the chemical properties of diffuse gas.
Before doing so, we explore further the underpinnings of our program and the 
determination of hydrogen column densities N(HI), N(\HH) and N(H)=N(HI)+2N(\HH) 
in diffuse molecular gas. We compare and merge observations of \HH\ and its 
proxies \hcop\ and CO along the various sightlines on which they have been 
observed in their respective wavebands. We ask whether we are getting the same 
idea of \HH\ formation in diffuse gas \citep{BelGod+20} from observations in the 
UV/optical and radio domains, and whether we are tracing the same chemical
behaviour in different environments.  To do so we need a common medium of 
exchange which we chiefly find in the reddening, \EBV, that is determined
photometrically toward stars and inferred from far infrared emission otherwise.

Results of observations in either domain in different directions can be
plotted against reddening whether determined spectroscopically toward a 
background star or using the all-sky maps of the FIR-determined optical 
reddening equivalent \citep{SchFin+98}. In turn, the reddening is commonly
used as a proxy for the total column density of H-nuclei in neutral
atoms and molecules, N(H) = N(HI) + 2N(\HH), offering the possibility of
determining the fraction of hydrogen in molecular form.  The atomic and
molecular  components 
of N(H) are observable along the sightlines used in UV absorption while the 
atomic component N(HI) is globally visible in \L21\ emission, including 
at \EBV\ $\la 0.1$ mag where N(\HH) is negligible. Somewhat surprisingly, 
it is found \cite{Lis14yEBV,Lis14xEBV} that the N(HI)/\EBV\ ratio at \L21\ 
at low \EBV\ (and perforce at high galactic latitudes), 
$ 8.3 \times 10^{21}~{\rm HI} \pcc$ (mag)$^{-1}$, 
is 40\% larger than the canonical value 
N(H)/\EBV\ $= 5.8 \times 10^{21}~{\rm H-nuclei} \pcc$ (mag)$^{-1}$  
determined by \cite{BohSav+78} toward stars. Here we show that a small
part of the difference can be ameliorated at low reddening using the 
most recent observations of N(HI) toward early-type stars, while the 40\% difference 
is confirmed overall and especially at \EBV\ $\ga$ 0.1 mag where N(\HH) 
is appreciable and the N(H)-\EBV\ relation toward stars is tighter and more robust.

The work is arranged 
as follows: Section 2 is a summary of the observational materials
discussed. Section 3 discusses variations in the UV-derived N(HI) and
N(\HH) with respect to \EBV\ and each other.  Section 4 merges the 
UV absorption and mm-wave measurements of \HH\ and CO. Section 5 is 
a summary.

%s2
\section{Datasets employed here}

\subsection{UV absorption}

\subsubsection{H$_2$}

We include previously published measurements of N(\HH) toward bright stars by 
\cite{BohSav+78}, \cite{BurFra+07}, \cite{SheRog+08}, \cite{RacSno+09} and 
\cite{ShuDan+21} and detections of N(\HH) in UV absorption against 34 
optically bright AGN at low reddening by \cite{GilShu+06}.

%1
\begin{figure*}
\includegraphics[height=8.2cm]{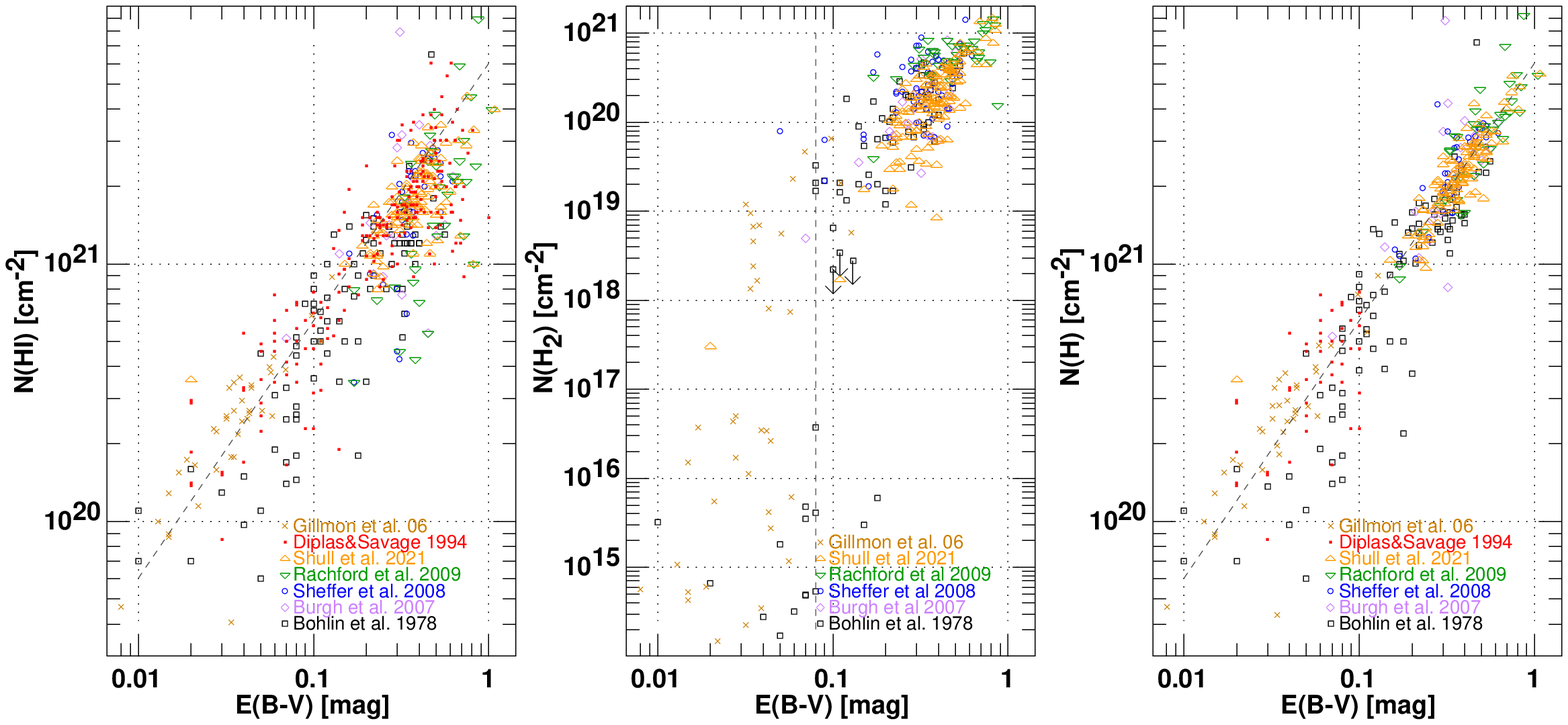}
\caption {N(HI), N(\HH) and N(H) = N(HI) + 2N(\HH) plotted against reddening as 
discussed in Section 2.3. Shown at 
left and right are dashed gray lines at N(HI)/\EBV\ and 
N(H)/\EBV\ $= 6\times 10^{21} \pcc$ mag$^{-1}$ respectively,
near the canonical value of \cite{BohSav+78}. Shown in the middle is a vertical 
dashed grey line at \EBV\ = 0.08 mag where N(\HH) canonically turns on in the 
diffuse ISM.  Most upper limits on N(\HH) are not physically meaningful and 
have been suppressed but a few are shown in the middle panel to indicate that 
lines of sight with low N(\HH) are occasionally seen even at \EBV\ $>$ 0.1 mag.
Also shown at right are values of N(HI) from \cite{DipSav94} at 
\EBV\ $\leq$ 0.1 mag where the molecular hydrogen fraction is small.
N(HI) from \cite{DipSav94} are only shown for sightlines lacking 
measured values of N(\HH).
The numbers of datapoints shown from left to right are 561, 370 and 381.}
\end{figure*}

%2
\begin{figure*}
\includegraphics[height=7.7cm]{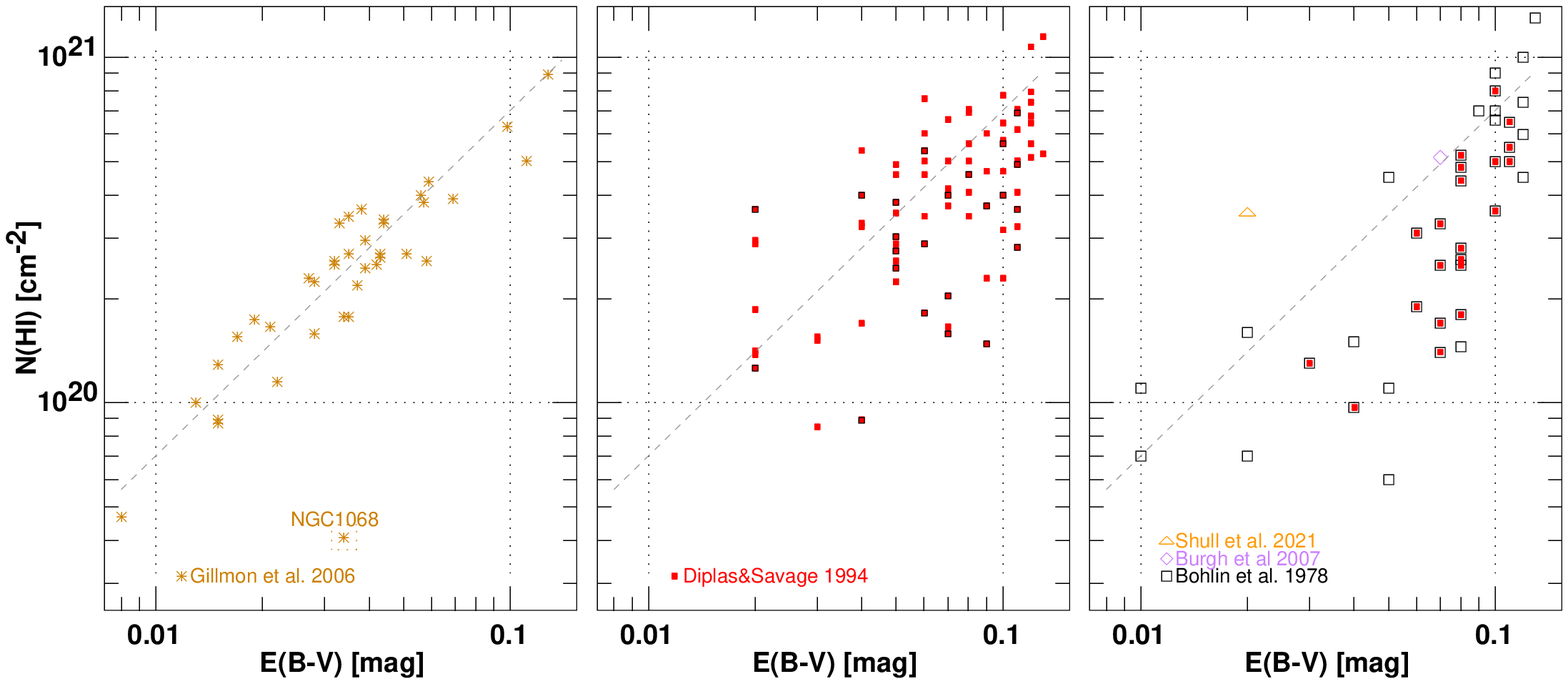}
\caption {N(HI) at low reddening. Left: from \cite{GilShu+06} using \EBV\ from  
\cite{SchFin+98} and N(HI) from \L21\ emission. Middle and right: from \cite{DipSav94} 
and \cite{BohSav+78}, respectively, using \EBV\ from stellar photometry and
N(HI) from stellar \Lya\ absorption. The dashed grey line in each panel is drawn 
at N(HI)/\EBV\ $= 7 \times10^{21}\pcc$ mag$^{-1}$, higher than the comparable lines 
drawn in Figure 1. Sightlines common to the samples of \cite{DipSav94} and
\cite{BohSav+78} are noted by black outlining in the middle panel and red fill 
at right: they have the same N(HI) and different \EBV\ in the two referencs
as remarked in Appendix C.  Results extracted from these samples are presented in 
Table 2.}
\end{figure*}

\subsubsection{HI}

The references cited in Section 2.1.1 reporting measurements of N(\HH) in 
UV absorption toward bright stars typically have complementary measured values 
of N(HI) measured in \Lya\ absorption, with occasional exceptions. 
\cite{BohSav+78} used the N(HI) values of \cite{SavDra+77}.
\cite{DipSav94} reported a survey of Ly-$\alpha$ measurements of N(HI) toward 
554 stars using IUE and this dataset was used to fill gaps in the entries for 
N(HI) in \cite{SheRog+08} and \cite{ShuDan+21} where possible. Such
sightlines are attributed to \cite{SheRog+08} or \cite{ShuDan+21}
as appropriate. The data of \cite{DipSav94} are used here along sightlines 
that do not have measured values of N(\HH).

Uniquely, \cite{GilShu+06} reported N(HI) measured in \L21\ emission, from 
the survey of \cite{WakKal+01} as noted in Section 2.2.4.  Their survey
of 45 sightlines included 38 with measured N(HI) and 7 with lower limits.

%Comparison of the measurements of N(HI) 
%between \cite{DipSav94} and other work having common sightlines showed a 
%difference of 0.2\% in the overall mean, with an rms standard deviation of 
%12\%.

\subsubsection{CO}

We include measurements of N(CO) in UV absorption toward bright stars from 
\cite{BurFra+07}, \cite{SonWel+07}, \cite{SheRog+07} and \cite{SheRog+08}, 
which we have considered in other recent references, especially 
\cite{Lis17CO,Lis20}.

\subsection{Radio frequency data}

\subsubsection{The H$_2$ surrogate \hcop}

We include previously reported values of N(\hcop) from \cite{LucLis96}, \cite{LisLuc00},
\cite{LisGer18} and \cite{LisGer+18} measured in absorption toward bright compact
extragalactic mm-wave continuum sources.  Also included and reported here for the first
time are  Cycle 6 ALMA results of the same kind in 33 directions toward ALMA phase
calibrators seen in the outer Galaxy, which resulted in 28 detections. As in 
\cite{LisGer18} and \cite{LisGer+18}, these are products of the regular pipeline, 
taken from the resultant datacubes at the peak of the continuum.  
The profile-integrated optical depth of \hcop\ in units of \kms\ is 
denoted by \Whcop.

\subsubsection{CO emission}

Reported here for the first time in Table 1 are line profile integrals \WCO\ from 
IRAM CO J=1-0 emission profiles taken in a frequency-switching mode in
August 2019 toward those 28 of the new 33 
outer Galaxy sightlines with detected \hcop. Five-point maps were made around
each direction, pointing toward the target and displaced by 1.2 half-power beamwidths 
(1.2x22\arcsec) in the four cardinal directions, and the profiles were co-added to 
produce the results shown here. The intensity scale is the usual main-beam intensity 
scale of the standard output.

Also reported here are older ARO Kitt Peak 12m results that had accumulated over 
the past several decades and were previously reported in \cite{LisPet+10} 
and \cite{LisPet12}.

%3
\begin{figure*}
\includegraphics[height=8.2cm]{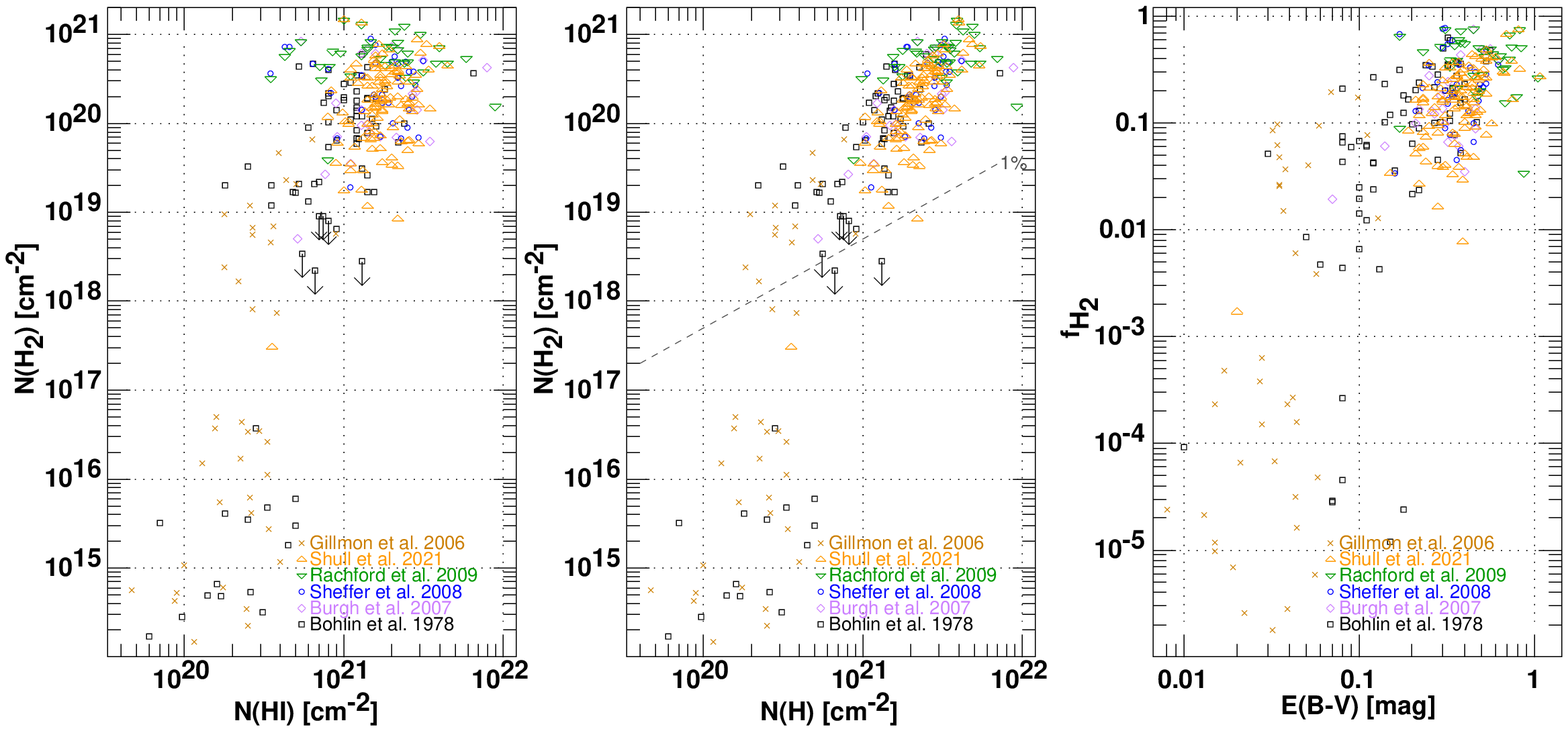}
\caption {N(\HH) plotted against N(HI) and N(H) at left and middle,
and the molecular hydrogen fraction plotted against reddening at right.
As in Figure 2 a few physically significant upper limits on N(\HH) are
shown to indicate that the appearance of sightlines with small N(\HH)
extends to somewhat higher N(HI) and N(H) than might otherwise be 
apparent. 318, 318 and 323 points representing measurements are shown
in the panels from left to right, respectively. The ensemble mean molecular
fraction for all measurements is $<$\fH2$>$ = 0.20 and \fH2\ = 
 $0.24\pm 0.16$ in a narrow interval around \EBV\ = 0.32 mag 
corresponding to \AV\ = 1 mag. The dashed line in the middle panel shows
the locus corresponding to \fH2\ = 0.01.}
\end{figure*}

\subsubsection{N(CO) measured in emission and/or absorption}

\cite{LisLuc98} measured N(CO) using a combination of profiles of the J=1=0 and J=2-1
lines measured in \cotw\ and \coth.  \cite{LisGer+18} extrapolated the methods of
\cite{LisLuc98} to measure N(CO) toward 6 directions in Chamaeleon from observations
of J=1-0 \cotw\ absorption.

\subsubsection{N(HI) measured in \L21\ emission}

In the absence of \Lya\ measurements in absorption toward AGN, \cite{GilShu+06} 
determined N(HI) in \L21\ emission from the high-velocity cloud survey of
\cite{WakKal+01}. 

\subsection{Reddening E(B-V)}

Observations of UV absorption toward bright stars report values for the optical 
reddening \EBV\ derived photometrically from stellar spectra. In the UV absorption
survey of \HH\ toward AGN \cite{GilShu+06} and for sightlines observed 
in CO and \hcop\ at mm-wavelengths, values of \EBV\ were taken from \cite{SchFin+98}.  
As noted in Section 3, recent measurements of N(HI) plotted against these two 
measures show very similar mean N(HI)/\EBV\ but very different systematic 
behaviour.

\subsection{Summary}

The means by which our references derive their values for \EBV, N(HI), N(\HH) and N(CO) 
are summarized in Table  4 in Appendix A.
The new CO and \hcop\ profiles used in this work are described in Appendix B and
shown in Figure 7.

%s3
\section{Relationships among UV-measured quantities}

We wish to use the reddening \EBV\ to infer N(H) along sightlines
studied at radio wavelengths and as a means of comparing observed and inferred 
results for N(\HH) between optical and radio wavebands, respectively. 
Ascertaining the N(H)/\EBV\ ratio 
took on new interest when it was realized that N(H)/\EBV\  could be 
substantially larger than the widely-used UV/optical-determined value 
$5.8 \times 10^{21}\pcc$ H-nuclei (mag)$^{-1}$\ \citep{BohSav+78} 
when sampled in \L21\ HI emission at Galactic latitudes above 20\degr\ 
\citep{Lis14yEBV,Lis14xEBV}.  Those references found a 40\% larger value
N(H)/\EBV\ $= 8.3 \times 10^{21}~{\rm H-nuclei} \pcc$ (mag)$^{-1}$. 
Other recent studies find or use ratios N(H)/\EBV\ =$7.7-9.4 \times 10^{21}
~{\rm H-nuclei} \pcc$ (mag)$^{-1}$ \citep{HenDra17,LenHen+17,LiTan+18}. 
\cite{HenDra20} use $8.8 \times 10^{21}~{\rm H-nuclei} \pcc$ (mag)$^{-1}$.

\subsection{N(HI)/E(B-V) and N(H)/E(B-V)}

The variations of N(H) and its constituents N(HI) and  N(\HH) determined in UV 
absorption are shown with respect to \EBV\ in Figure 1. Shown at left and right 
is a line representing a column density/reddening ratio N(HI)/\EBV\ or N(H)/\EBV\ 
= $6 \times 10^{21}~{\rm H-nuclei} \pcc$ (mag)$^{-1}$,
ie the old standard value.  Comparing the panels at left and right for N(HI) 
and N(H), it is striking how summing the atomic and molecular contributions 
tightens the column density-reddening relationship at \EBV\ $\ga 0.1$, removing 
for instance most of the 
very large scatter in N(HI) around \EBV\ $\approx\ 0.3-0.4$ mag in the
strongly molecular sample of \cite{RacSno+09}. Moreover, the old standard
ratio N(H)/\EBV\ $= 5.8 \times 10^{21}~{\rm H-nuclei} \pcc$ (mag)$^{-1}$ is a good
representation of the more strongly-molecular data at \EBV\ $\ga$ 0.2 mag,
as noted recently by \cite{ShuDan+21}.  

The regression lines for all the data and for \EBV\ $\geq$ 0.2 mag are the same 
in Figure 1 at right, with power-law slopes of 1.0 and 
N(H)/\EBV\ = $6.2 \times 10^{21}~{\rm H-nuclei} \pcc$ (mag)\m1. 
This is only 6\% higher than the old standard value 
and 2\% larger than the quantity calculated by \cite{ShuDan+21}
$<$N(\HH)/\EBV$> = (6.07\pm1.01)\times 10^{21}\pcc$ (mag)\m1 . 

%That being said, it is 
%still something of a lower limit because, as seen at right in Figure 1, there 
%are a substantial number of sightlines lacking measurement of N(\HH) at 
%\EBV\ $>$ 0.2 mag in the work of \cite{DipSav94} where 
%N(HI)/\EBV\ $> 6\times 10^{21}~{\rm H-nuclei} \pcc$ (mag)$^{-1}$. 
%In general, N(\HH) is not negligible in such cases.

The absence of a tighter column density-reddening relationship in atomic and
weakly-molecular gas at \EBV $\la 0.2$ mag in Figure 1 is something of a puzzle, 
given the lack of competition for H-nuclei between atomic and molecular hydrogen. 
At high Galactic latitude and small reddening where \HH\ is unimportant, 
the column density-reddening relationship determined in \L21\ HI emission
is tight down to \EBV\ = 0.01 mag meaning that the contribution of H\p\
is not important. 

To explore this, Figure 2 shows expanded plots of N(HI) against reddening at 
\EBV\ $\la$ 0.1 mag for the different datasets.  The data of \cite{GilShu+06} 
using the IR-determined 
reddening and N(HI) measured in \L21\ emission are shown separately at left:
The other datasets use stellar reddening and N(HI) measured in Ly-$\alpha$
absorption. The IUE data of \cite{DipSav94} are shown in the middle while the rest 
of the UV absorption line measurements of N(HI), almost all from the
original $Copernicus$ sample of \cite{BohSav+78}, are shown at right.  There is 
a variety of different behaviour and the statistics of the data at 
\EBV\ $\leq 0.08$ mag are summarized in Table 2.

The N(HI)-\EBV\ relationship is only tight in the highly-ordered dataset of 
\cite{GilShu+06} that does not use stellar measurements. The data appear
to depart from a simple linear relation at higher \EBV\ where N(\HH) may 
be significant.  The data of 
\cite{DipSav94} have  a similar ensemble mean $<$N(HI)$>$/$<$\EBV$>$ to that 
of \cite{GilShu+06} at slightly higher $<$\EBV$>$ (Table 2) but considerably 
more scatter. The scatter must in part arise from the limited precision of 
the quoted reddening, but also from random error in the determination of \EBV\ 
from stellar spectra. Comparison of measurements of N(HI) and N(HI)/\EBV\ along 
sightlines observed in common between \cite{BohSav+78} and \cite{DipSav94} 
in Appendix C leaves little doubt that this is the case.

The two newer datasets of \cite{DipSav94} and \cite{GilShu+06} comprise 80\% 
of the measurements below \EBV\ = 0.08 mag and (see Table 2) have a mean 
$<$N(HI)$>$/$<$\EBV$>$ = $6.8 \times 10^{21}~{\rm H-nuclei} \pcc$ (mag)$^{-1}$.
In a log sense this spans about half of the difference between the 
values 5.8 and 8.3 $\times 10^{21}~{\rm H-nuclei} \pcc$ (mag)$^{-1}$.
The dataset of \cite{BohSav+78} has 40\% lower ensemble mean 
$<$N(HI)$>$/$<$\EBV$>$ = $4.1\times 10^{21}\pcc$ (mag)$^{-1}$ than the newer 
data. A more detailed comparison is given in Appendix C.

We summarize these results by concluding first that N(HI)/\EBV\ measured 
at low extinction, and N(H)/\EBV\ overall, are smaller than the mean N(HI)/\EBV\ 
determined from \L21\ emission above latitude 20\degr\ for the Galaxy at large.  
As shown in Table 2, a new value of $<$N(H)$>$/\EBV\ derived from the four times 
larger dataset shown in Figure 1 at all \EBV\ would 
be $<$N(H)$>$/$<$\EBV$>$ = $6.2 \times 10^{21}~{\rm H-nuclei} \pcc$ (mag)$^{-1}$, 
an increase of only 7\% over that of \cite{BohSav+78}. At the same time, the mean 
gas/reddening ratio N(HI)/\EBV\ at \EBV\ $\leq$ 0.08 mag in Table 2,
$6.8 \times 10^{21}~{\rm H-nuclei} \pcc$ (mag)$^{-1}$, is larger
than the mean over all reddening but still smaller than the comparable 
high-latitude Galactic average.

\subsection{N(H$_2$) vs reddening}

The behaviour of N(\HH) with \EBV\ is shown in the middle panel of Figure 1.
Subsequent to the original survey of \cite{BohSav+78}, observers of \HH\ toward 
bright stars 
%\footnote{ perhaps mindful of the expected behaviour of time allocation 
%committees leery of wasting too much effort on negative results}
studied sightlines with appreciable N(\HH)/N(H). This has the somewhat unfortunate
consequence that the rise of N(\HH) at small \EBV\ is largely documented in the 
sightlines of \cite{GilShu+06} toward AGN that represent different conditions 
and/or a different reddening measurement method from the bulk of the data overall. 
As can be seen in Figure 1, a few sightlines with small N(\HH)
persist up to \EBV\ $\la 0.2$ mag. but in general, sightlines with 
N(\HH) $<< 10^{19}\pccc$ at \EBV\ $>$ 0.08 mag are rare. 

%4
\begin{figure*}
\includegraphics[height=9.8cm]{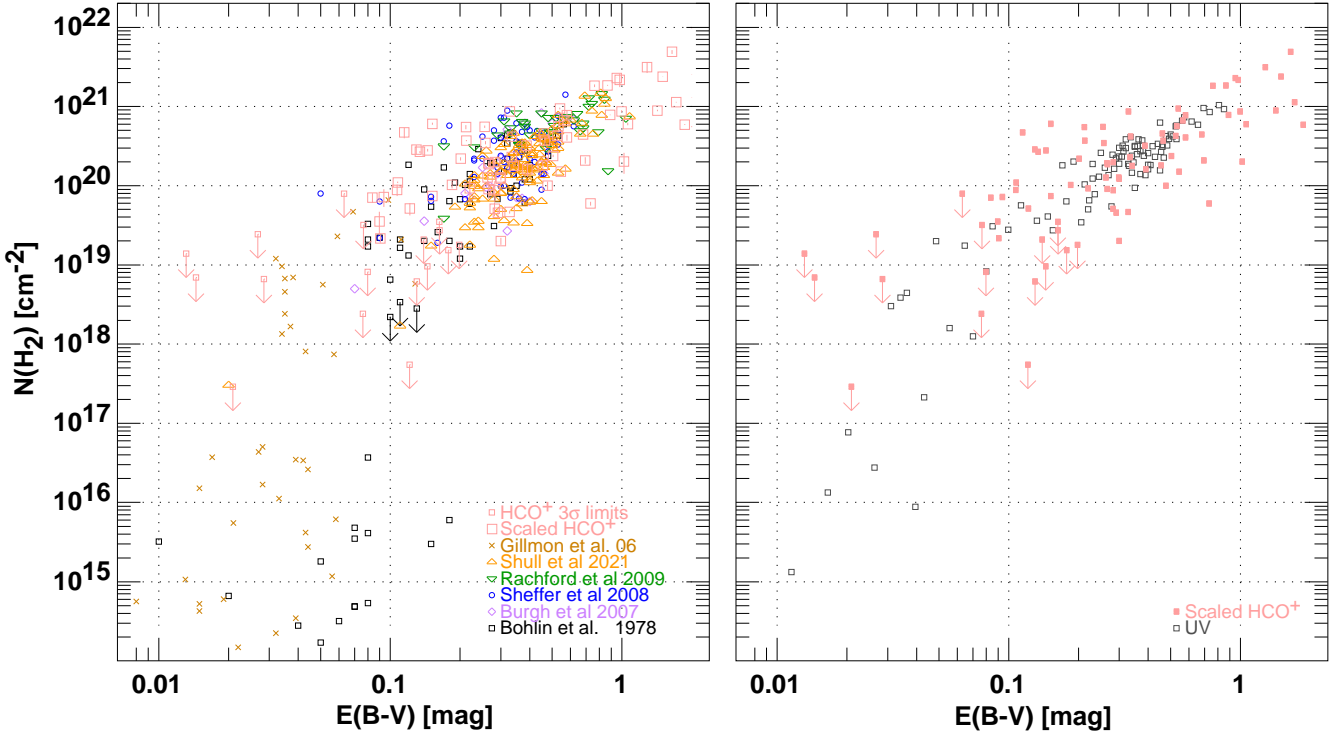}
\caption {N(\HH) plotted against reddening, using the data from Figure 1 
and merging the UV and mm-wave absorption results by scaling 
N(\HH) = N(\hcop)$/3\times10^{-9}$. At right the UV datasets were merged 
and sorted in \EBV, then averaged horizontally and vertically in sequential 
non-overlapping groups of four to reduce the scatter and facilitate comparison.
See Table 3 for a quantitative comparison.}
\end{figure*}

%5
\begin{figure*}
\includegraphics[height=9.0cm]{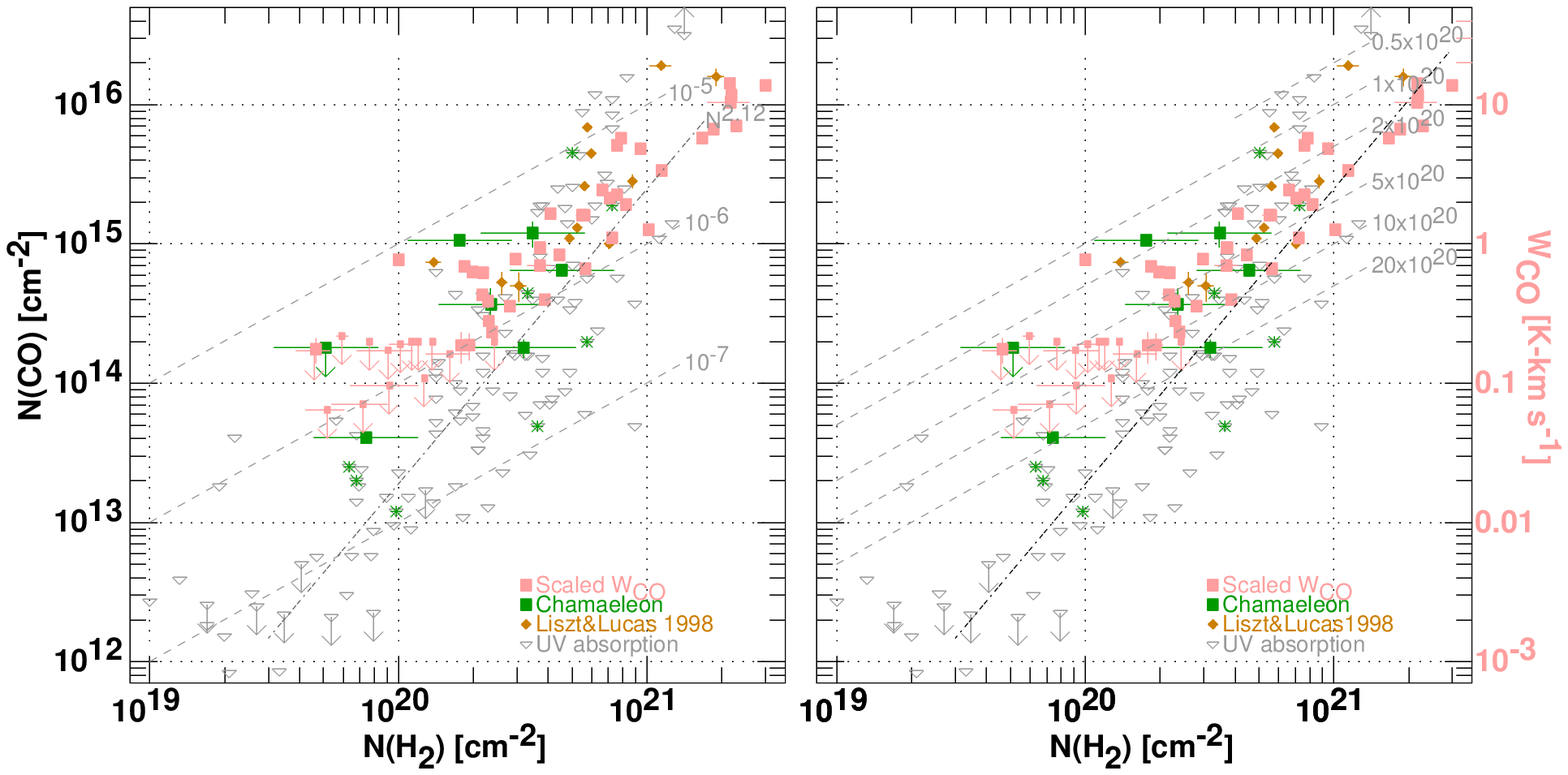}
\caption {N(CO) and \WCO\ plotted against N(\HH) merging radio and optical data 
with two scalings: N(\HH) = N(\hcop)$/3\times10^{-9}$ and 
N(CO) = \WCO\ $\times 10^{15} \pcc$ (K-\kms)$^{-1}$. N(CO) and N(\HH) values 
derived
from UV absorption are plotted as a background of gray downward-pointing
open triangles.  Symbols shown in green are results from our earlier 
observations in Chamaeleon \citep{LisGer+19} with filled rectangles 
representing N(CO) derived in $\lambda2.6$mm absorption and green asterisks 
marking the positions of 8 Chamaeleon stars  in the wider stellar UV absorption 
sample. Gray dashed lines represent the fractional abundance N(CO)/N(\HH) at 
left and the CO-\HH\ conversion factor N(\HH)/\WCO\ at right. The dash-dotted
line with power-law slope 2.12 is a regression fit to a portion of the data as 
discussed in Section 4.2}
\end{figure*}

%6
\begin{figure}
\includegraphics[height=7.7cm]{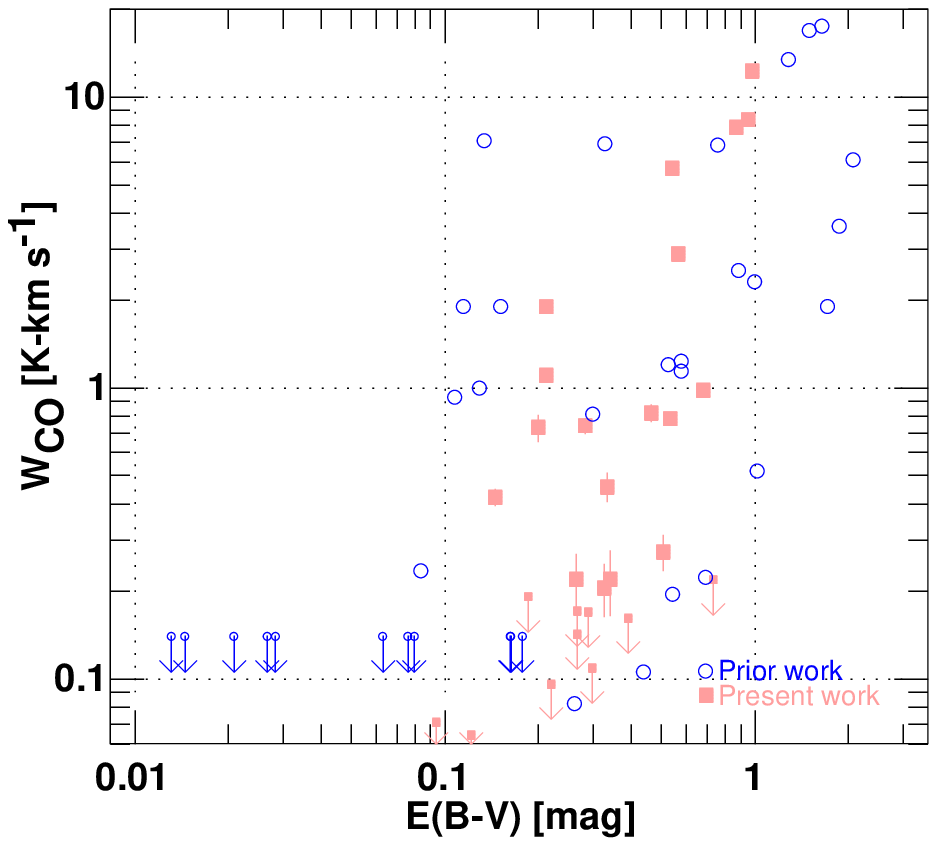}
\caption {CO emission \WCO\ plotted against reddening \EBV.}
\end{figure}

\subsection{Variation of N(H$_2$) and the molecular fraction with N(HI) and N(H)}

For completeness and to facilitate consideration of the molecular fraction, 
Figure 3 shows the variation of N(\HH) with respect to N(HI) and N(H).
The variation of the molecular fraction 
\fH2\ = 2N(\HH)/(N(HI) + 2N(\HH)) 
with N(H) is implicit in the middle panel of Figure 3 
and the variation of \fH2\ with \EBV\ is shown at right there. \cite{BelGod+20} 
recently used  a version of the middle panel in Figure 3, with data compiled by 
\cite{GudBub+12} lacking the recent sample of \cite{ShuDan+21}, to show that
the statistics of \HH\ formation can be reproduced in a simulation of the 
turbulent ISM. 

The ensemble mean molecular fraction for all the data in Figure 3 is 
$<$\fH2$>$ = 0.20.  The mean molecular fraction in a narrow interval
\AV\ = 0.95 - 1.05 mag, with \AV\ = 3.1 \EBV, is only slightly larger
0.24$\pm$0.16. 

For reference, recall that a Spitzer ``standard HI cloud'' \cite{Spi78} has 
N(H) $= 4\times 10^{20} \pcc$ and \EBV\ = 0.07 mag, and according to 
Figures 1 and 3, such a cloud will have a molecular fraction \fH2\ of 
order 1-10\%. The equilibrium value of \fH2\ at the center of a quiescent 
HI cloud model resembling the Spitzer Standard Cloud is about one quarter 
\citep{Lis07}.  

%s4
\section{Merging UV absorption and mm-wave measurements}

To merge the UV/optical and radio datasets we consider two aspects; using \hcop\
as a proxy for \HH\ and using N(CO) and \WCO\ interchangeably as proxies for each 
other. If these are possible, the fractional CO abundances and CO-\HH\ conversion 
factors derived in the two regimes may be compared.

\subsection{HCO\p\ considered as a proxy for H$_2$}

Shown in Figure 4 are the UV/optical absorption measurements of N(\HH) plotted
against \EBV\ as in Figure 3, with mm-wave measurements put on the same scale
using N(\HH) = N(\hcop)$/3\times10^{-9}$. At right in Figure 4, the cloud of UV 
absorption measurements was collapsed to facilitate comparison: The individual UV 
datasets were merged, the resultant dataset was sorted in \EBV\ and the data were 
averaged horizontally and vertically in sequential non-overlapping groups of four. 

On the scale of Figure 4 there is no offset between the radio and UV datasets: 
the behaviour of N(\HH) vs \EBV\ is the same.  Table 3 shows that the ensemble 
means $<$N(\HH)$>$/$<$\EBV$>$ measured in the UV/optical and radio/IR are within 
about 0.1 dex (25\%) and the properties of the radio data are indistiguinshable
from those of the three most recent UV absorption studies, especially \cite{RacSno+09} 
that has comparable $<$\EBV$>$. Within the scatter, which is large for both N(\HH) 
and N(\hcop) vs. \EBV, \hcop\ is a reliable proxy for \HH\ with the assumption 
that N(\HH) = N(\hcop)$/3\times10^{-9}$. Demonstrating the coincidence of the 
UV and mm-wave determinations of N(\HH) is the most important result
of this work and was never a foregone conclusion.  It is perhaps surprising,
considering that the UV absorption measurements sample gas toward relatively rare, 
early-type stars that are surrounded by natal gas and have a significant
influence on their immediate environment.  
Environmental effects are more clearly manifested when CO is considered,

\subsection{W$_{CO}$ and N(CO) as mutual proxies}

In Figure 5 we plot CO column densities and integrated brightness \WCO\ 
against N(\HH), using the usual fractional abundance N(\hcop)/N(\HH)
$= 3\times 10^{-9}$ and the additional conversion 
\WCO\ = N(CO)$/10^{15}\pcc$ in units of K-\kms\ as in \cite{Lis17CO}.
This relationship was evident  when the J=1-0 and J=2-1 transitions
of CO were observed in emission and absorption by \cite{LisLuc98},
and was seen to apply even when the CO optical depths were substantial. 
It can be understood as the result of weak (strongly sub-thermal) 
rotational excitation at the kinetic temperatures typical of diffuse 
molecular gas \citep{GolKwa74,Lis07CO}.

In both panels of Figure 5, 113 UV absorption measurements of N(CO) and 
N(\HH) are shown as a greyed background. The mm-wave
measurements take several forms as noted in Section 2.2. Results based solely 
on J=1-0 emission measurements (\WCO) are plotted as pink filled rectangles 
with smaller symbols representing 3$\sigma$ upper limits.  Six measurements
of N(CO) toward mm-wave background sources in Chamaeleon based on J=1-0 
absorption are shown as filled green rectangles, and green asterisks denote 
eight stellar lines of sight where N(CO) and N(\HH) were measured in UV 
absorption toward stars in Chamaeleon \citep{LisGer+18}. The N(CO) values of 
\cite{LisLuc98} based on J=1-0 and J=2-1 absorption and emission measurements 
are shown as brown filled diamonds.

At left in Figure 5 the vertical scale is N(CO) and shaded dashed lines represent 
the fractional abundance N(CO)/N(\HH). The vertical scale at right 
is in units of integrated brightness, K-\kms, and the shaded dashed 
lines in the right panel represent N(\HH)/\WCO, the CO-\HH\ conversion factor.  
Shown in both panels for reference as a shaded dash-dotted line is the power-law 
regression line fit to the UV absorption data at 
N(\HH) $\geq 7\times 10^{19}\pcc$ that has N(CO) $\propto$ N(\HH)$^{2.12}$.

The radio data inhabit those portions of the plane above the regression
line for the UV data and are much more tightly correlated with N(\HH) than
are the UV data. The slope of a power-law regression line through
the radio data, \WCO\ = 0.0030 K-\kms\ (N(\HH)$/10^{19}\pcc)^{1.556}$. 
is less steep.  Unlike in the UV, there are no sightlines observed
in the radio with N(\HH) $\ga 2\times10^{20}\pcc$ and N(CO)/N(\HH) $< 10^{-6}$.
This difference is common to all the radio datasets, representing not only the
scaled \WCO\ measurements in pink but the CO column densities  directly determined 
at radio wavelengths using CO absorption \citep{LisLuc98,LisGer+18}. 

Moving the radio values of N(CO) to the right in Figure 5 to align them with the 
regression line of the UV absorption measurements would require an increase in 
N(\HH) by a factor $\approx 1.6 = 0.2$ dex, ie a decrease in the N(\hcop)/N(\HH) 
relative abundance of the same size. This is at the edge of the range believed to 
apply to empirical determination of N(\hcop)/N(\HH) as estimated by 
\cite{GerLis+19}.  It would introduce a noticeable displacement between the 
UV and radio values for N(\HH)/\EBV\ whose good agreement was discussed in 
Section 4.1.

Moving the radio datapoints vertically downward would require a factor three 
smaller N(CO)/\WCO\ ratio that can be excluded on physical grounds if the CO 
emission originates in diffuse molecular gas. As discussed by \cite{Lis17CO} 
the conversion N(CO)/\WCO\ $= 10^{15}\pcc/$K-\kms\ is a broad local 
minimum with respect 
to the number density in calculations of the rotational excitation. N(CO)/\WCO\ 
will increase for n(H) $\la 50 \pccc$ to compensate for the small amounts of energy 
being deposited in CO by collisions. At very high n(H) $\ga 1000 \pccc$ collisions 
will increasingly populate rotational energy levels above J=1, also increasing 
N(CO)/\WCO.

The difference in behaviour between the UV and radio data in Figure 5 stands in 
contrast to the similar behaviour of N(\HH) vs \EBV\ in Figure 4
where both the UV and radio datasets contain sightlines with small N(\HH)/\EBV, 
even with appreciable reddening. But CO observed at radio wavelengths also differs 
from that observed in UV absorption in having more fractionation from C\p\ 
isotope exchange \citep{Lis17CO}. The differences in the behaviour of the radio and 
UV datasets in Figure 5 are real, not artifacts of the scalings and proxy 
relationships.  

As discussed by \cite{SheRog+08}, the CO column density for given N(H) and 
N(\HH) strongly depends on the local UV radiation field, with larger CO columns 
for a lower radiation field. As CO is shielded by dust, by \HH\ and by itself 
for larger N(CO) \citep{VisVan+09}, the somewhat larger N(CO) and N(\coth)/N(CO) 
of the radio samples may be due to small variations in the sampled environments 
due to differences in the selection criteria of the background sources and the 
intrinsic variability of conditions in the diffuse interstellar gas
\citep{HuSte+21}.

%
%\section{Inferences}%
%
%Drawing generalities is an imperfect art and CO emission with 
%\WCO\ $\approx$ 1 K-\kms\ is sometimes seen at \EBV\ as small as 
%0.1 mag \cite{LisPet12} but CO in Figure 5 is generally detected 
%%in mm-wave emission at the  level 
%\WCO\ $\ga$ 0.1 K-\kms\ when 
%N(\HH) $\ga 2\times 10^{20}\pcc$, 
%N(CO)/N(\HH) $\ga 10^{-6}\pcc$ and 
%N(\HH)/N(CO) $\la 10 \times 10^{20} \HH\ \pcc$.  
%CO was detectable at the typical CO survey limit \WCO\ = 1 K-\kms\
%for N(\HH) $\ga 4\times 10^{20}\pcc$.
%In this regime the CO relative abundance N(CO)/N(\HH) and the 
%CO-\HH\ conversion factor N(\HH)/\WCO\ are mutual proxies, related 
%by the conversion factor between CO column density and brightness
%\WCO\ = N(CO)$/10^{15}\pcc$ K-\kms.

\subsection{The CO-H$_2$ conversion factor}

As indicated in Figure 5, equivalence of \WCO\ and N(CO) in diffuse molecular 
gas creates an equivalence between the CO relative abundance N(CO)/N(\HH) and the
CO-\HH\ conversion factor $\alpha$ = N(\HH)/\WCO. 
%The sky map of CO emission and molecular gas is the map of CO chemistry.  
There is a wide variation in $\alpha$ observed at radio wavelengths and an 
even larger implied variation in the UV absorption data. Nevertheless,
the ensemble mean $<$N(\HH)$>$/$<$\WCO)$>$ $= 2.4 \times 10^{20}$\HH\ (K-\kms)$^{-1}$
or $= 2.1 \times 10^{20}$ \HH (K-\kms)$^{-1}$ when only the data with \WCO\ $\ge$ 1
K-\kms\ is considered.  This is the same result reached on other grounds without
relying on the relative abundance of \hcop\ by \cite{LisPet+10}: the mean
CO-\HH\ conversion factor in the diffuse molecular gas is just the usual 
standard value\footnote{\cite{LadDam20} recently revised the local value of the 
CO-\HH\ conversion factor upward by a factor 1.8 to
N(\HH)/\WCO\ = $3.6 \times 10^{20} \HH \pcc$ (K-\kms)$^{-1}$.
This is close to the originally-suggested but subsequently-eclipsed
value N(\HH)/\WCO\ = $5 \times 10^{20}~\HH \pcc$ (K-\kms)$^{-1}$
of \cite{Lis82}.}.
A physical explanation offered by \cite{LisPet+10} for this coincidence noted 
that the small relative abundance of CO, here $<$N(CO)/N(\HH)$> = 4\times 10^{-6}$, 
is compensated by a very high brightness per CO molecule
\WCO/N(CO) = 1 K-\kms\ $\times$ N(CO)/$10^{15}\pcc$ in the J=1-0 line,
resulting from the strongly sub-thermal excitation. 

\subsection{The molecular fraction of gas observed in CO emission}

Mean molecular fractions $<$\fH2$>$ = 0.2 - 0.24 were derived in Section 3.3 
for the lines of sight sampled in \HH\ in UV absorption, as shown in Table 3. 
This is the same value suggested by the original $Copernicus$ sample after 
correction for a bias toward low mean line of sight density and low molecular 
fraction \citep{BohSav+78}. The molecular fraction of the gas observed in
\hcop\ absorption is higher, 0.25-0.35 
for N(H)/\EBV\ $= 6-8.3\times10^{21}\pcc$ mag$^{-1}$, as also shown in Table 3.   

The CO-\HH\ conversion factor can  be used to constrain the mean molecular 
fraction in the diffuse molecular gas where CO emission was observed.
Expressing N(H) = $\beta$\EBV\ and N(\HH)= \fH2 N(H)/2, we have 
N(\HH)/\WCO\ = ($\beta$\fH2/2) \EBV/\WCO. The variation of \WCO\ with \EBV\ 
is shown in Figure 6:  For the 58 sightlines observed in CO emission 
at \EBV\ $\geq$ 0.07 mag, taking upper limits on \WCO\ at their 3$\sigma$ 
values, one has $<$\WCO$>$ = 2.492 K-\kms\ and $<$\EBV$>$ = 0.529 mag.
With $\beta = 6.2 \times 10^{21}~{\rm H-nuclei} \pcc$ (mag)$^{-1}$ the
implied mean molecular fraction in the CO-detected gas is 0.34, near
the upper end of the range in Table 4 based on \EBV\ and N(\HH) alone.
This higher mean molecular fraction in gas observed away from bright 
stars could account for the occurence of the radio CO
data in Figure 5 above the regression line for the CO observed in 
UV absorption.

%t1
\begin{table*}
\caption{Line of sight properties and new results for \hcop\ and CO }
{
\small
\begin{tabular}{lcccccccccc}
\hline
Source & RA(J2000)&Dec(J2000)& $l$ & $b$ &\EBV & flux& \Whcop $^2$ &$\sigma$\Whcop & \WCO & $\sigma$\WCO \\  
      &  hh.mmsss&dd.mmsss&degrees & degrees &mag &Jy & \kms & \kms & K-\kms& K-\kms \\  
\hline
J0203+1134$^1$& 2.03464& 11.34492&149.6826&-47.4992&0.144&0.126&$\leq$0.263&0.088& & \\
J0209+1352& 2.09357& 13.52045&150.1800&-44.8290&0.094&0.223& 0.20&0.050&$\leq$0.071&0.024\\
J0211+1051& 2.11132& 10.51348&152.5781&-47.3674&0.145&0.462& 0.76&0.029& 0.36&0.025\\
J0213+1820& 2.13105& 18.20255&148.7289&-40.4014&0.130&0.093&$\leq$0.345&0.115& & \\
J0231+1322& 2.31459& 13.22547&157.0917&-42.7380&0.121&0.430& 0.14&0.025&$\leq$0.064&0.021\\
J0242+1742& 2.42243& 17.42589&157.0180&-37.7033&0.077&0.227&$\leq$0.151&0.050& & \\
J0252+1718& 2.52077& 17.18427&159.7420&-36.7885&0.220&0.172& 0.25&0.077&$\leq$0.096&0.032\\
J0325+2224& 3.25368& 22.24004&163.6700&-28.0213&0.213&1.162& 1.01&0.017& 0.94&0.051\\
J0329+3510& 3.29154& 35.10060&155.9152&-17.4042&0.267&0.570& 0.51&0.032&$\leq$0.143&0.048\\
J0329+2756& 3.29577& 27.56155&160.7030&-23.0743&0.198&0.158&$\leq$0.193&0.064& & \\
J0334+0800& 3.34533&  8.00144&177.2396&-37.0871&0.391&0.150& 0.44&0.088&$\leq$0.162&0.054\\
J0336+3218& 3.36301& 32.18293&158.9998&-18.7650&0.733&1.689& 0.16&0.009&$\leq$0.219&0.073\\
J0356+2903& 3.56085& 29.03423&164.6120&-18.4927&0.212&0.139& 1.50&0.090& 1.62&0.042\\
J0357+2319& 3.57216& 23.19538&169.0302&-22.4661&0.185&0.224& 0.28&0.027&$\leq$0.192&0.064\\
J0400+0550& 4.00117&  5.50431&184.2710&-33.7266&0.266&0.159& 0.25&0.063&$\leq$0.172&0.057\\
J0401+0413& 4.01199&  4.13344&186.0261&-34.4947&0.341&0.405& 0.49&0.021& 0.19&0.048\\
J0403+2600& 4.03056& 26.00015&168.0260&-19.6483&0.201&0.331& 0.60&0.029& 0.62&0.067\\
J0406+0637& 4.06343&  6.37150&184.7075&-32.0009&0.283&0.264& 0.54&0.051& 0.63&0.040\\
J0407+0742& 4.07291&  7.42075&183.8723&-31.1558&0.265&0.387& 0.53&0.031& 0.19&0.042\\
J0426+0518& 4.26366&  5.18199&189.3631&-28.7705&0.291&0.516& 0.12&0.020&$\leq$0.170&0.057\\
J0426+2327& 4.26557& 23.27396&173.8881&-17.4457&0.539&0.304& 2.57&0.057& 4.84&0.045\\
J0427+0457& 4.27476&  4.57083&189.8857&-28.7306&0.335&0.414& 0.62&0.024& 0.39&0.045\\
J0437+2037& 4.31038& 20.37343&176.8096&-18.5565&0.532&0.217& 1.54&0.073& 0.67&0.026\\
J0431+1731& 4.31574& 17.31358&179.4942&-20.3579&0.464&0.104& 1.01&0.110& 0.70&0.049\\
J0433+0521& 4.33111&  5.21156&190.3730&-27.3967&0.298&4.911& 0.35&0.003&$\leq$0.109&0.036\\
J0437+2940& 4.37044& 29.40138&170.5818&-11.6609&0.979&0.059& 5.92&1.138&10.42&0.027\\
J0438+3004& 4.38049& 30.04455&170.4116&-11.2283&0.952&0.689& 6.25&0.038& 7.11&0.026\\
J0439+3045& 4.39178& 30.45076&170.0655&-10.5913&0.867&0.195& 5.05&0.082& 6.69&0.027\\
J0440+1437& 4.40211& 14.37570&183.2538&-20.5438&0.681&0.337& 1.21&0.031& 0.83&0.029\\
J0445+0715& 4.45014&  7.15539&190.4535&-23.8898&0.121&0.275&$\leq$0.083&0.028& & \\
J0449+1121& 4.49077& 11.21286&187.4274&-20.7365&0.504&0.521& 0.65&0.022& 0.23&0.033\\
J0502+1338& 5.02332& 13.38110&187.4143&-16.7456&0.564&0.271& 1.81&0.059& 2.46&0.031\\
J0510+1800& 5.10024& 18.00416&184.7304&-12.7895&0.328&2.411& 0.13&0.005& 0.17&0.036\\
\hline
\end{tabular}
\\
$^1$ The coordinates of J0203+1134 should be read as 2$^{\rm h}$03$^{\rm m}$46.4$^{\rm s}$,11\degr 34\arcmin 49.2\arcsec  \\
$^2$ \Whcop\ is the integrated optical depth of \hcop\ in units of \kms \\
}
\end{table*}

%t2
\begin{table*}
\caption{\EBV\ and N(HI) at \EBV\ $\leq 0.08$ mag }
{
\small
\begin{tabular}{lcccc}
\hline
sample & n & $<$\EBV $>$ & $<$N(HI)$>$ &  $<$N(HI)$>$/$<$\EBV$>$ \\  
        &        &    mag    & $\pcc$        & $\pcc$ mag$^{-1}$     \\
\hline
\cite{BohSav+78} etc. $^1$ & 26 & 0.0569  & 2.332$\times10^{20}$ & 4.10$\times10^{21}$ \\
\cite{DipSav94} & 50 & 0.0526  & 3.563$\times10^{20}$ & 6.77$\times10^{21}$ \\
\cite{GilShu+06}$^2$ & 34 & 0.0348 & 2.409$\times10^{20}$ & 6.92$\times10^{21}$ \\
Two above & 84 & 0.0454 & 3.096$\times10^{20}$ & 6.82$\times10^{21}$ \\
All \EBV\ $\leq 0.08$ mag & 111 & 0.0477  & 2.949$\times10^{20}$ & 6.18$\times10^{21}$ \\
\hline
\end{tabular}
\\
$^1$ Results from \cite{BohSav+78}, \cite{BurFra+07} and \cite{ShuDan+21} in Figure 2 at right
\\
$^2$ Sightlines with measured N(HI), ignoring toward NGC 1068 (see Figure 2)
\\
}
\end{table*}

%t3
\begin{table*}
\caption{Survey mean and ensemble parameters }
{
\small
\begin{tabular}{lrccccc}
\hline
Source for N(\HH) &n$^1$ & $<$\EBV$>$  & $<$N(\HH)$>$  & $<$N(\HH)$>$/$<$\EBV$>$ & $<$\fH2$>$ & $<$N(H)$>$/$<$\EBV$>$\\
       &  &  mag   & log $\pcc$ &log $\pcc$ mag$^{-1}$  &   &    $10^{21}\pcc$ mag$^{-1}$\\       
\hline
\cite{BohSav+78} & 76 &0.212 &20.040 &20.710 &0.177 &5.41\\
\cite{GilShu+06} &38 &0.041 &18.753 &20.141 &0.040 &6.88\\
\cite{BurFra+07} &24 &0.349 &20.458 &20.915 &0.225 &7.30\\
\cite{SheRog+08}&64 &0.339 &20.497 & 20.967 &0.267 &6.66\\
\cite{RacSno+09}&38 &0.519 &20.791 &21.075 &0.370 &6.43\\
\cite{ShuDan+21}&112$^2$&0.396 &20.412 &20.814 &0.215 & 5.95\\
\hline
All UV  & 378 & 0.323& 20.40 &20.89 &0.213 & 6.13 \\
%All UV-Gillmon  & 340 & 0.355& 20.44&20.89 & 0.232 & 6.02 \\
\hline
Radio unflagged$^3$  & 88 &0.563 &20.772 & 21.02 & 0.253-0.350$^4$ &\\ 
Radio flagged$^4$ &  71 &0.674 &20.862 &21.03 &0.260-0.359$^5$ &\\ 
\hline
\end{tabular}
\\
$^1$ Numbers of targets with measured N(\HH) \\
$^2$ 112 targets had measured N(HI) and N(\HH) \\
$^3$ Upper limits are taken at the $3\sigma$ value \\
$^4$ Sightlines with upper limits are ignored \\
$^5$ For N(H)/$<$\EBV$>$ = $6 - 8.3 \times 10^{21}\pcc$ mag$^{-1}$
\\
}
\end{table*}

%s5
\section{Summary}

We began by noting the complementary nature of UV/optical and radio observations of
diffuse, partially molecular interstellar gas.  The rich polyatomic chemistry of diffuse
molecular gas is only apparent in absorption at radio wavelengths but the column densities 
of atomic and molecular hydrogen  determined in UV absorption provide a characterization of 
the host gas that is not matched in the radio domain.  Observations in common of 
CH, CO and OH provide the rudiments of a bridge between the radio and UV/optical domains
and determinations of reddening \EBV\ are available through different means, which we
use throughout this work. 

Prior observations indicated that the relative abundance of the widely observable 
polyatomic ion \hcop\ is constant in diffuse molecular gas, with 
N(\hcop)/N(\HH) $= 3\times 10^{-9}$.  In that case, \EBV\ and 
N(\hcop), together with the usual assumed total hydrogen/reddening ratio N(H)/\EBV, give 
the relative abundances of molecules with respect to \HH\ and the fraction of H-nuclei in 
\HH, \fH2\ = 2N(\HH)/N(H). In turn, mm-wave observations of CO in emission determine the 
CO-\HH\ conversion factor in the diffuse molecular gas.  

Our focus, then, was on three aspects of the observations: determining N(HI)/\EBV\ 
and N(H)= N(HI)+2N(\HH))/\EBV (Figures 1-3 and Tables 2-3 and 5); comparing 
N(\HH) and \EBV\ measured at UV and radio wavelengths; and comparing observed and implied 
column densities N(CO) and integrated J=1-0 brightnesses \WCO\ of carbon monoxide.

We first discussed the N(H)/\EBV\ ratio that was originally determined in optical 
UV/absorption by $Copernicus$ to be 
N(H) $= 5.8 \times 10^{21}~{\rm H-nuclei} \pcc$ (mag)$^{-1}$ but subsequently was 
found to be globally larger, $8.3 \times 10^{21}~{\rm H-nuclei} \pcc$ (mag)$^{-1}$, 
at high galactic latitude when deriving N(HI) from \L21\ emission and \EBV\ from scaled
far-IR emission.  The tension between these values is not resolved
by the vastly increased number of UV-absorption measurements undertaken subsequent to
$Copernicus$.  Overall and especially at \EBV\ $\ga 0.1$ mag where the molecular 
fraction is appreciable and the plot of N(H) vs \EBV\ wraps itself tightly about the
regression line in Figure 1, we find that  
N(H) $= 6.2 \times 10^{21}~{\rm H-nuclei} \pcc$ (mag)$^{-1}$ is a good average for
sightlines with N(\HH) measured in UV absorption.

That said, $<$N(HI)$>$/$<$\EBV$>$ and the scatter in N(HI) vs \EBV\ are larger at 
\EBV\ $<$ 0.08 mag (Figure 1) where there is no competition for H-nuclei between 
atomic and molecular hydrogen. In Figure 2 and Tables 2 and 5 we remarked several 
properties of the UV absorption measurements at \EBV\ $\leq 0.08$ mag of N(HI) 
subsequent to $Copernicus$:  they have an HI/reddening ratio 
$<$N(HI)$>$/$<$\EBV$>$ = $6.8 \times 10^{21}~{\rm H-nuclei} \pcc$ (mag)$^{-1}$ 
that is much larger than in the $Copernicus$ sample
($4.1 \times 10^{21}~{\rm H-nuclei} \pcc$ (mag)$^{-1}$ in Table 2)
but is equal to the ratio seen toward optically bright AGN when N(HI) 
is measured in \L21\ emission and \EBV\ is inferred from far-IR emission 
($6.9 \times 10^{21}~{\rm H-nuclei} \pcc$ (mag)$^{-1}$).  Similarity of the
mean values of $<$N(HI)$>$/$<$\EBV$>$ in these post-$Copernicus$ samples suggests
that scatter in the plot of stellar measurements of N(HI) vs \EBV\ is due to 
errors in the stellar photometry that is demonstrated in Appendix C and Figure 8.
 Subdividing the stellar absorption
data into subsamples of common and unshared sightlines at small and large
reddening in Appendix C showed no calibration differences in measured N(HI)
but notable differences in \EBV\ for shared sightlines at low \EBV, and
very different explanations for disparities in N(HI)/\EBV\ in the same sense. 

% Including the $Copernicus$ measurements, the full dataset has 
%$<$N(HI)$>$/$<$\EBV$>$ = $6.6 \times 10^{21}~{\rm H-nuclei} \pcc$ (mag)$^{-1}$
%at \EBV\ $\leq$ 0.08 magnitude, showing that the HI/reddening ratio is larger
%for sightlines lacking \HH.
 
In Figure 4 we merged the radio and UV determinations of N(\HH) and jointly plotted 
them against \EBV.  They coincide and have comparable scatter when \HH\ is detected. 
They also do not contradict each other at low reddening where N(\HH) is small and 
\hcop\ is not detected (Table 3).  Thus, radio and optical data provide the same view of the 
HI$\rightarrow$\HH\ transition in diffuse gas and a numerical comparison is given in 
Table 3. This commensurability is gratifying and lends confidence to the use of 
N(\hcop)/N(\HH) $= 3\times 10^{-9}$ along sightlines observed at radio wavelengths. It 
has the consequence that if N(\hcop) is measured and N(\HH) is derived indirectly by 
some other means, the implied relative abundance N(\hcop)/N(\HH) should be consistent 
with $ 3\times 10^{-9}$. Possible use cases are when \EBV\ is scaled to N(H) and a value 
of N(HI) is subtracted to determine N(\HH) or when 
\WCO\ is scaled to N(\HH) and subtracted from N(H) to find N(HI) and \fH2. 

We merged the UV and radio determinations of N(CO) and plotted them against N(\HH) 
in  Figure 5, testing the equivalence of \WCO\ and N(CO), 
\WCO\ = N(CO)$/10^{15}\pcc$ K-\kms. This was first observed when N(CO) was derived 
by observing CO emission and absorption in the lower rotation levels, and it applies 
even to optically thick CO lines over the observed range of 
N(CO) $\la 10^{16}\pcc$.  It can be understood by 
realizing that photons eventually escape after scattering in a medium
where the excitation of CO is weak (significantly sub-thermal),  even when 
the J=1-0 line is optically thick.

Here a difference was found: The radio and UV data overlap but the radio data 
inhabit only the upper portions of the region in the N(CO)-N(\HH) plane that
are occupied by the UV data.  Unlike in the UV, sightlines are not present 
in the radio dataset with N(\HH) $\ga 2\times10^{20}\pcc$ and 
N(CO) $< 10^{14}\pcc$ or N(CO)/N(\HH) $< 10^{-6}$. In Section 4.4 we noted 
that the molecular fraction appears to be higher in the gas observed in 
CO emission and \hcop\ absorption, ~0.34 vs ~0.22.

Radio measurements of \WCO\ 
are much more tightly correlated with N(\hcop) than are N(CO) and N(\HH) 
measured in UV absorption.  Overall, there is a power-law relation between 
\WCO\ and N(\HH) measured in the radio domain 
\WCO\ = 0.0030 K-\kms\ (N(\HH)$/10^{19}\pcc)^{1.556}$. 

Variations in N(CO)/N(\HH) (Figure 5) and N(\HH)/\EBV\ (Figure 1) drive 
varations in N(\HH)/\WCO\ (aka the CO-\HH\ conversion factor) and \WCO/\EBV\ (Figure 6).
Nonetheless, the overall ensemble mean for the radio data discussed in Section 4.3,
$<$N(\HH)$>$/$<$\WCO$>$ = $2.1-2.4 \times 10^{20} \HH \pcc$ (K-\kms)$^{-1}$, 
is like those in common use in dense and dark gas, 
\WCO/N(\HH) $= 2-4 \times 10^{20} \HH \pcc$ (K-\kms)$^{-1}$.  
In statistical terms, this similarity in the ensemble mean arises because 
so much of the CO emission is contributed by lines of sight at higher \EBV\
where the relative CO abundance is higher.
In terms of the physics, similarity of the CO-\HH\ conversion factors in 
diffuse and fully molecular dense gas arises from compensating influences of 
the CO chemistry and excitation: The smaller relative abundance N(CO)/N(\HH) in 
low density diffuse molecular gas is compensated by a high brightness in the 
J=1-0 line per CO molecule (\WCO/N(CO)) resulting from the strongly sub-thermal 
excitation.

\begin{acknowledgments}
  The National Radio Astronomy Observatory is a facility of the National
  Science Foundation operated under contract by Associated  Universities, Inc.
  This work was supported in part by the Programme National 
  “Physique et Chimie du Milieu Interstellaire” (PCMI) of CNRS/INSU with 
  INC/INP co-funded by CEA and CNES. 
  
  This work is  based in part on observations carried out under project number 
  003-19 with the IRAM 30m telescope. IRAM is supported by INSU/CNRS (France), 
  MPG (Germany) and IGN (Spain). The assistance of the IRAM staff and,
  especially, help from Axel García Rodríguez, is very much appreciated.
  
  This paper makes use of the following ALMA data: ADS/JAO.ALMA\#2018.1.00115.S,
  ADS/JAO.ALMA\#2017.1.00120.S and \\
  ADS/JAO.ALMA\#2016.1.00714.S. 
  ALMA is a partnership of ESO (representing its member states), NSF (USA) 
  and NINS (Japan), together with NRC (Canada), NSC and ASIAA (Taiwan), 
  and KASI (Republic of Korea), in cooperation with the Republic of Chile. 
  The Joint ALMA Observatory is operated by ESO, AUI/NRAO and NAOJ.

  We thank Isabelle Grenier for comments on earlier versions of this work
 and we thank the referee, Michael Shull, for many helpful remarks.

\end{acknowledgments}

\appendix

%\restartappendixnumbering

\section{Sources for column densities and reddening}

Shown in Table 4 are the sources and methods by which reddenings and 
column densities were derived for references cited here for their
measurement of N(HI), N(\HH) and N(CO).

%t4
\begin{table}
\caption{Sources for Column Densities and Reddening }
{
\small
\begin{tabular}{lcccc}
\hline
Reference & \EBV & N(HI) & N(\HH)& N(CO) \\
\hline
\cite{BurFra+07} &  Stellar Photometry &  UV Absorption &  UV Absorption & UV Absorption  \\
\cite{SheRog+08}& $\prime\prime$ & $\prime\prime$ & $\prime\prime$ & $\prime\prime$  \\
\cite{RacSno+09}& $\prime\prime$ & $\prime\prime$ & $\prime\prime$ & $\prime\prime$  \\
\cite{ShuDan+21} & $\prime\prime$ & $\prime\prime$ & $\prime\prime$ & $\prime\prime$  \\
\cite{BohSav+78} &  $\prime\prime$ & $\prime\prime$ & $\prime\prime$ & NA \\
\cite{DipSav94} & $\prime\prime$ & $\prime\prime$ & NA & NA  \\
\cite{GilShu+06} & \cite{SchFin+98} & \L21 emission & UV Absorption & NA  \\
\cite{LisLuc98} & $\prime\prime$ & NA & Scaled \hcop & mm-wave emission \& absorption \\
\cite{LisGer+19} & $\prime\prime$ & NA & $\prime\prime$ & mm-wave absorption  \\
\hline
\end{tabular}
\\
}
\end{table}

%AppendixB
\section{Spectra of \hcop\ and CO}

%7
\begin{figure*}
\includegraphics[height=21cm]{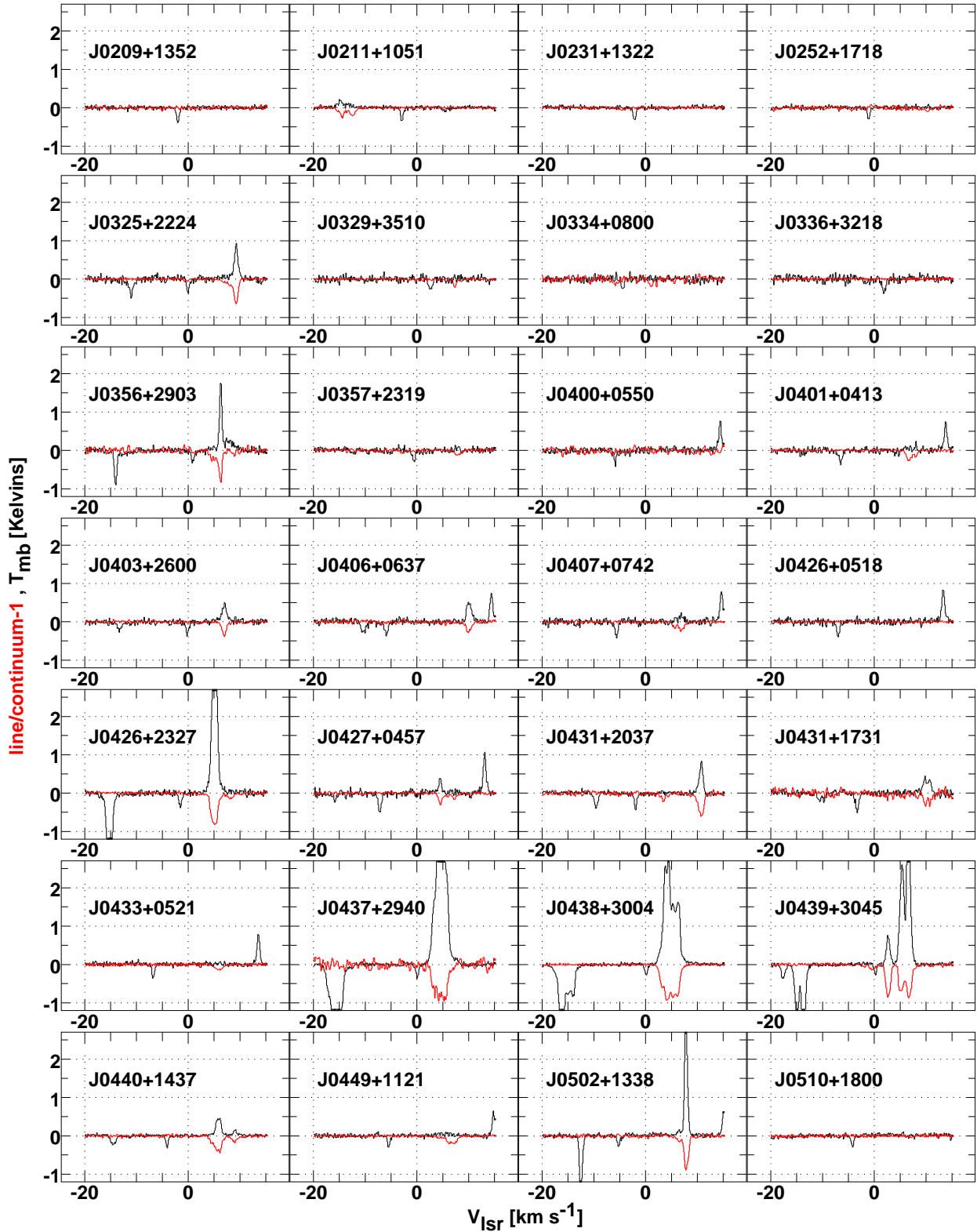}
 \caption{Spectra of  \hcop\ absorption (red) and CO emission (black) from the Galactic 
 anticenter sample. See Appendix B for an explanation of the appearance of the 
 frequency-switched IRAM 30m emission spectra.}
\end{figure*}

Spectra of \hcop\ and CO in the 28 directions with detected \hcop\ absorption are shown 
in Figure 7. The \hcop\ absorption is the negative-going signal shown in red.  The 
black histogram shows the CO emission. The as-delivered frequency-switched IRAM 30m CO 
spectra were, unfortunately, already folded in frequency, preventing a separation of
the phases of the frequency switching cycle using the methods of \cite{Lis97FS}.
Only the positive-going signal coincident with \hcop\ absorption is interstellar,
signals at the 0.5 - 1 K level not coincident with \hcop\ is telluric.

%AppendixC
\section{Stellar determinations of N(HI) and E(B-V) at low  E(B-V)}

The discussion in Section 3, along with Figure 2 and the numerical results
in Table 2, showed a significant disparity between the 
N(HI)/\EBV\ ratios measured toward stars by \cite{DipSav94} and \cite{BohSav+78} at 
\EBV\ $\le 0.08$ mag where \fH2\ is negligible. Meanwhile, \cite{GilShu+06} arrived 
at the same N(HI)/\EBV\  as \cite{DipSav94} while deriving \EBV\ from \cite{SchFin+98} 
and N(HI) from \L21\ emission, with much smaller scatter in the plot of N(HI) vs \EBV\ 
(Figure 2). In Section 3.1 we ascribed the larger scatter in the stellar measurements 
of N(HI)/\EBV\ to error in the photometric determination of \EBV\ that
is consistent with having the same mean.

Shown in Figure 8  are the measurements of 
N(HI) and N(HI)/\EBV\ for sightlines in common between \cite{BohSav+78}
and \cite{DipSav94}: 15 and 13 stars respectively for which \EBV\ $\le 0.08$ 
mag in either dataset and 63 stars in common overall. Comparison of the tightly 
correlated plots of column density at left with the plots of column density/reddening 
ratio at right clearly show the scatter introduced by differences in the stellar reddening.

Numerical results 
are abstracted in Table 5 for increasingly larger samples from both datasets.
Results for the 63 common sightlines should serve as a fiducial and they show 
agreement in the mean \EBV\ and N(HI) at the 1\% level. 
This overall agreement renders the results at low \EBV\ somewhat confounding. 
{\it Considering only the data at \EBV\ $\leq 0.08$ mag,} the much larger N(HI)/\EBV\ 
in \cite{DipSav94} along common sightlines arises from a difference in \EBV\ 
with nearly equal N(HI), while for the whole low reddening sample the difference 
arises from much larger N(HI) in \cite{DipSav94} with nearly equal \EBV.
The common sightlines have small N(HI)/\EBV\ even in \cite{DipSav94}.

%tA2
\begin{table}
\caption{\EBV\ and N(HI) at \EBV\ $\leq 0.08$ mag and in general}
{
\small
\begin{tabular}{lccccc}
\hline
Authors & sub-sample & n  & $<$\EBV $>$ & $<$N(HI)$>$ &  $<$N(HI)$>$/$<$\EBV$>$ \\  
        &        &   &   mag    & $10^{20}\pcc$   & $10^{21}\pcc$ mag$^{-1}$ \\
\hline
\cite{BohSav+78} & Common \EBV\ $\le 0.08$ mag  & 15& 0.069(0.015)   &2.69(1.25)  & 3.90 \\
\cite{DipSav94} &  ''                           & 13& 0.054(0.015)   &2.80(1.28)  & 5.20 \\
\cite{BohSav+78} & Unique \EBV\ $\le 0.08$ mag  & 11& 0.041(0.025)   &1.85(1.96)  &3.83  \\
\cite{DipSav94} &  ''                           & 37& 0.052(0.020)   &3.83(1.75)  & 7.34 \\
\cite{BohSav+78} etc & All \EBV\ $\le 0.08$ mag$^1$ & 26 & 0.057(0.024)  & 2.33(1.64) & 4.10 \\
\cite{DipSav94}  & '' & 50                &0.053(0.019)  & 3.56(1.70) & 6.77 \\
\cite{BohSav+78} & All common, any \EBV    & 63& 0.212(0.130)   &9.55(8.73)  & 4.50 \\
\cite{DipSav94} &    ''                   & 63& 0.210(0.130)   &9.55(7.01)  & 4.55 \\
\cite{BohSav+78}& All in dataset            & 90  & 0.198(0.131) &8.56(7.92)  & 4.32  \\
\cite{DipSav94} &    ''                   & 388 & 0.312(0.180) &15.33(7.80) & 4.91 \\
\hline
\end{tabular}
\\
$^1$ From Table 2
\\
}
\end{table}

%8
\begin{figure*}
\includegraphics[height=14cm]{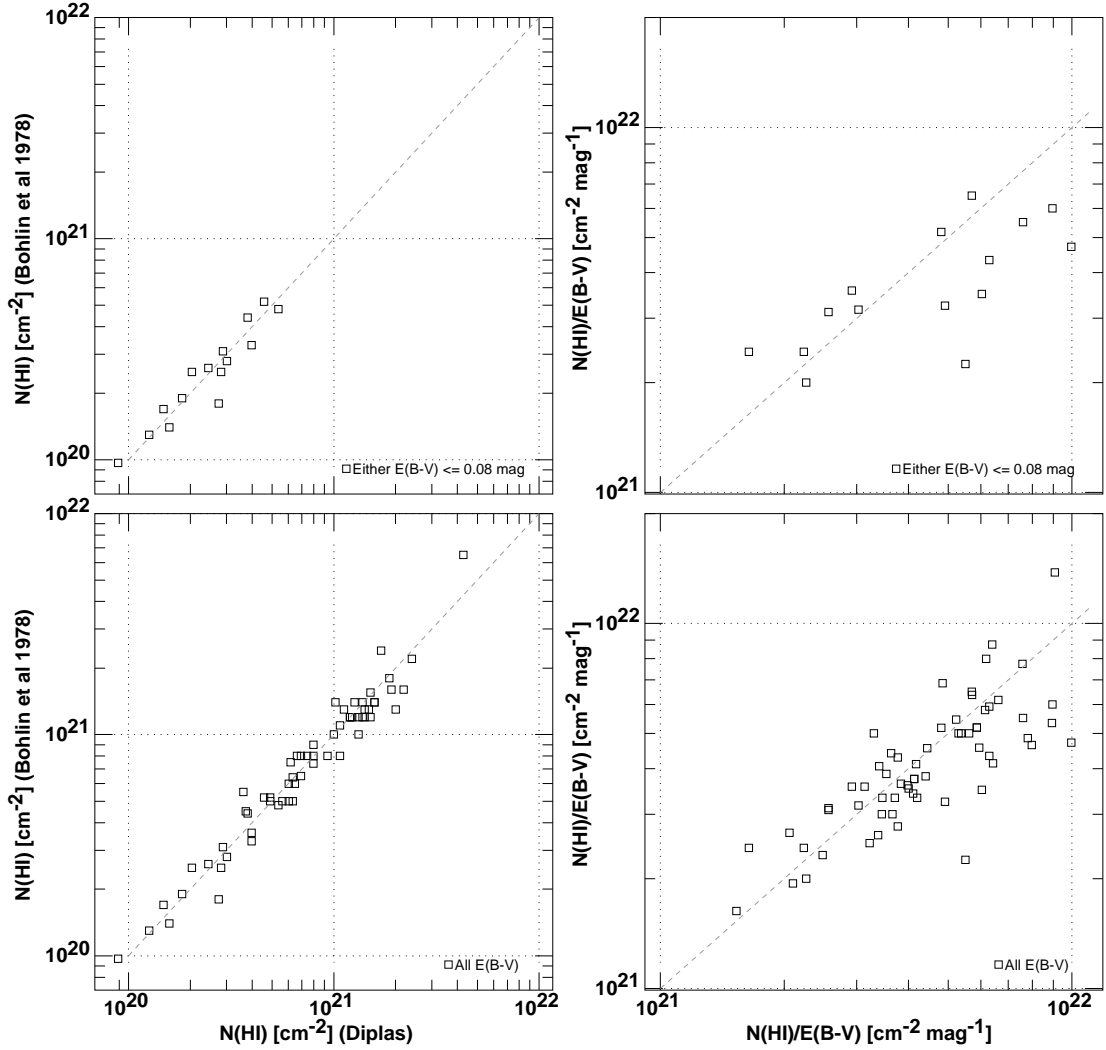}
\caption {Comparison of N(HI) and N(HI)/\EBV\ from \cite{DipSav94} plotted horizontally 
and \cite{BohSav+78} plotted vertically in all panels . Top: 15 common sightlines for which 
\EBV\ $\leq 0.08$ in either or both datasets. Bottom: all (63) common sightlines.  The gray 
dashed line in each panel is the locus of equal values, not a regression line.} 
\end{figure*}

 %%%%%%%%%%%%%%%%%%%%%

% \bibliographystyle{apj}
% \bibliography{mnemonic,absorption}

 %%%%%%%%%%%%%%%%%%%%

 \end{document}